\documentclass[sigconf, authorversion]{acmart}

\AtBeginDocument{%
  \providecommand\BibTeX{{%
    \normalfont B\kern-0.5em{\scshape i\kern-0.25em b}\kern-0.8em\TeX}}}

\copyrightyear{2024} 
\acmYear{2024} 
\setcopyright{rightsretained} 
\acmConference[CUI '24]{ACM Conversational User Interfaces 2024}{July 8--10, 2024}{Luxembourg, Luxembourg}
\acmBooktitle{ACM Conversational User Interfaces 2024 (CUI '24), July 8--10, 2024, Luxembourg, Luxembourg}
\acmDOI{10.1145/3640794.3665563}
\acmISBN{979-8-4007-0511-3/24/07}

\begin{document}

\title[Embedding Large Language Models into Extended Reality]{Embedding Large Language Models into Extended Reality: Opportunities and Challenges for Inclusion, Engagement, and Privacy}

\author{Efe Bozkir}
\affiliation{%
 \department{Human-Centered Technologies for Learning}
 \institution{Technical University of Munich}
 \city{Munich}
 \country{Germany}}
\email{efe.bozkir@tum.de}
\affiliation{%
 \department{Human-Computer Interaction}
 \institution{University of T{\"u}bingen}
 \city{T{\"u}bingen}
 \country{Germany}}
\email{efe.bozkir@uni-tuebingen.de}
\orcid{0000-0002-4594-4318}

\author{S{\"u}leyman {\"O}zdel}
\affiliation{%
 \department{Human-Centered Technologies for Learning}
 \institution{Technical University of Munich}
 \city{Munich}
 \country{Germany}}
\email{ozdelsuleyman@tum.de}
\orcid{0000-0002-3390-6154}

\author{Ka Hei Carrie Lau}
\affiliation{%
 \department{Human-Centered Technologies for Learning}
 \institution{Technical University of Munich}
 \city{Munich}
 \country{Germany}}
\email{carrie.lau@tum.de}
\orcid{0009-0005-8838-3230}

\author{Mengdi Wang}
\affiliation{%
 \department{Human-Centered Technologies for Learning}
 \institution{Technical University of Munich}
 \city{Munich}
 \country{Germany}}
\email{mengdi.wang@tum.de}
\orcid{0000-0002-5254-737X}

\author{Hong Gao}
\affiliation{%
 \department{Human-Centered Technologies for Learning}
 \institution{Technical University of Munich}
 \city{Munich}
 \country{Germany}}
\email{hong.gao@tum.de}
\orcid{0000-0003-3934-433X}

\author{Enkelejda Kasneci}
\affiliation{%
 \department{Human-Centered Technologies for Learning}
  \institution{Technical University of Munich}
  \city{Munich}
  \country{Germany}}
\email{enkelejda.kasneci@tum.de}
\orcid{0000-0003-3146-4484}

\renewcommand{\shortauthors}{Bozkir, et al.}

\begin{abstract}
Advances in artificial intelligence and human-computer interaction will likely lead to extended reality (XR) becoming pervasive. While XR can provide users with interactive, engaging, and immersive experiences, non-player characters are often utilized in pre-scripted and conventional ways. This paper argues for using large language models (LLMs) in XR by embedding them in avatars or as narratives to facilitate inclusion through prompt engineering and fine-tuning the LLMs. We argue that this inclusion will promote diversity for XR use. Furthermore, the versatile conversational capabilities of LLMs will likely increase engagement in XR, helping XR become ubiquitous. Lastly, we speculate that combining the information provided to LLM-powered spaces by users and the biometric data obtained might lead to novel privacy invasions. While exploring potential privacy breaches, examining user privacy concerns and preferences is also essential. Therefore, despite challenges, LLM-powered XR is a promising area with several opportunities. 
\end{abstract}




\maketitle

\section{Introduction}
With research and development in computer graphics, artificial intelligence (AI), and human-computer interaction fields, virtual, mixed, and augmented reality (VR/MR/AR) and head-mounted displays (HMDs) have started to become pervasive in everyday life. Especially with big tech companies like Apple and Meta bringing their HMDs into the market for vast use (e.g., Apple Vision Pro, Meta Quest 3), it is likely that HMDs will soon become similar to today's smartphones, smartwatches, or tablets. Yet, with HMDs, users often experience highly immersive virtual scenes, making their interaction and societal aspects different from the other devices. 

Extended reality, covering the wide spectrum of VR, AR, and MR, has been used for different purposes in various domains, such as education~\cite{gao_2021_chi, bozkir_2021_ieeevr, Hayes2021}, medicine~\cite{Hombeck_2022_ieeevr, ar_vr_in_medicine_2021, peters2018mixed}, entertainment~\cite{bates_1992, kodama_2017_3dui, HUNG2021101757}, transportation~\cite{Faria_2021, 8797758, McGill2020}, and business~\cite{MEINER2020219, Rejeb_2021, gillopez_2023}. One of the advantages of XR and HMDs, especially in understanding human behaviors and interactions, is the possibility of obtaining fine-grained motion sensor data, such as eye- and head-tracking~\cite{glancable_ar_2020}, possibly with controlled stimuli~\cite{bozkir_thesis_2022}. This allows a better understanding of how humans behave and perceive in XR and provides an opportunity for real-time and adaptive user support. Combining such real-time and adaptive interaction experiences with high presence, immersion, and sociality levels might lead to positive user experiences, motivating users to use XR and HMDs frequently. 

Despite several advantages, one of the issues in terms of sociality with virtual and mixed spaces that include non-player characters (NPCs), even those powered by AI and machine learning, is the limited conversational capabilities of the characters. These characters are often designed as pre-programmed agents~\cite{hong_2022_animations_vrst} or trained for particular use with relevant training data~\cite{Kastanis_Slater_2012}. However, these cannot be utilized flexibly for open-ended conversations or open-world settings. Furthermore, when the audience changes (e.g., adults to children), conversational scripts or trained models must be updated almost from scratch, requiring significant human labor to create scripts or label training data. 

Embedding large language models (LLMs) as chat agents into XR spaces can significantly solve these issues and help these spaces provide more inclusive and engaging experiences. LLMs are trained with a good portion of the Internet and fed with a diverse set of text data hence, they can converse about various topics. In addition, the possibility of fine-tuning such models with use-case-specific small datasets and prompt engineering helps them be utilized for a broader range of tasks. Especially with ChatGPT being publicly introduced in 2022~\cite{chatgpt_public_2022}, a wide audience has also observed how powerful LLMs are and the human-like text they can generate. Possibilities and opportunities have been emphasized for different domains, such as education~\cite{KASNECI2023102274} and medicine~\cite{Singhal2023}. However, they have not been addressed and studied in depth for XR yet, except from few works mentioning their potential as conversation agents in XR~\cite{virtual_aivantage_2023, xr_ai_ws_2023, Waisberg2023} or demonstrating their abilities to produce and edit objects, and scenes in MR~\cite{delatorre2023llmr}. One of the reasons might be the uncommon use of textual interaction in XR compared to auditory and visual. 

Despite the higher likelihood of auditory and visual information use in XR than text, it is trivial to build automated processing pipelines, with speech-to-text~\cite{pratap2023scaling}, large language~\cite{touvron2023llama}, and text-to-speech~\cite{pratap2023scaling} models, to process audio data through LLMs for XR. Additionally, with the emergence of multimodal LLMs, such as GPT-4~\cite{openai2023gpt4v} and Gemini~\cite{Deepmind2023}, the use of LLMs within XR spaces to facilitate conversations will be more intuitive than pre-programming the avatars or using conventional AI techniques. Therefore, this paper argues for embedding LLMs in XR as virtual avatars or narratives. This process will promote more inclusive, diverse, and engaging experiences with three main implications, as in the following. 

\begin{itemize}
  \item We first argue that embedding LLMs into XR as NPCs or narratives with different prompting strategies and model fine-tuning will support designing more inclusive settings. This will also support diversity and equity in XR. 
  \item We state that with the multifaceted conversational abilities of LLMs, LLM-powered spaces will facilitate more engaging XR experiences for users, a step towards pervasive XR. 
  \item Increased engagement with NPCs and XR likely means that users will provide more information about themselves. We state that combining this information with the biometric sensor data from XR will cause novel privacy invasions. These possible invasions should be investigated, along with user privacy concerns, to enable user-centered XR spaces. 
\end{itemize}

\section{Related Works}
As we argue for using LLMs in XR by embedding them into NPCs or treating them as narratives to facilitate inclusion, diversity, and engagement by taking ethical aspects, especially privacy, into account, we summarize previous literature in two folds. We first discuss the recent works on LLMs in Section~\ref{lbl_subsec_llms_rw}. Then, we provide previous literature on inclusion, diversity, equity, and privacy in XR in Section~\ref{lbl_subsec_xr_rw}. 

\subsection{Large Language Models}
\label{lbl_subsec_llms_rw}
Large language models are specific types of artificial neural networks trained with massive amounts of data, usually by scrapping a significant portion of the Internet, and they can generate human-like text and facilitate natural conversations~\cite{KASNECI2023102274}. One of the reasons that LLMs are successful in natural language processing (NLP) tasks is the transformer architecture and the self-attention mechanism~\cite{vaswani2017attention}. Particularly, a transformer is a self-supervised encoder-decoder model, and the self-attention mechanism assumes that some words are more related to each other than others and operates according to this assumption to find the relationships. Due to massive amounts of training data and these methods, LLMs can operate in various tasks very well, especially for tasks whose data is widely available online. 

Prevalent examples of LLMs are OpenAI's GPT-3~\cite{Floridi2020}, Meta's LLaMa~\cite{touvron2023llama}, Google's PaLM~\cite{chowdhery2022palm, anil2023palm}, and more recently multimodal LLMs such as GPT-4~\cite{openai2023gpt4v} and DeepMind's Gemini~\cite{Deepmind2023}. In addition to these general-purpose pre-trained LLMs, to facilitate chat scenarios, researchers further iterated the LLMs to achieve naturalistic interactions by fine-tuning them to align them in the conversations~\cite{touvron2023llama2}. Different techniques also exist for facilitating this with other steps, such as aligning language models to follow instructions~\cite{ouyang2022training}. Yet, fine-tuning these models with high-quality labels is often necessary to facilitate specific uses and personalization. 

Several works fine-tuned pre-trained LLMs for particular purposes, including software bug fixing~\cite{jin2023inferfix}, dialogue summarization in customer services~\cite{finetuned_llms_dialogue_summarization_2023}, understanding patients' needs and providing informed advice~\cite{li2023chatdoctor}, and legal knowledge understanding~\cite{fei2023lawbench}. These works showed evidence that fine-tuned models enhance task performance, hinting at better personalization. Apart from fine-tuning, it is also known that different prompting techniques can significantly help in getting more tailored responses, even without fine-tuning~\cite{prompt_engineering_openai, wei2023chainofthought, besta2023graph, yao2023tree}. It is also possible to create LLM-based mechanisms that access external knowledge sources~\cite{NEURIPS2020_RAG}, which can facilitate adaptive solutions. While the users can partially carry out these methods, automating these processes based on the user characteristics in the backend is essential for LLM use in XR. Apart from the scientific literature, OpenAI launching a GPT app store~\cite{openai_gpt_store} and smart NPCs powered by OpenAI or users' language models being integrated into Unreal Engine~\cite{unreal_npc_openai} are other examples that LLMs can be utilized in XR. Yet, despite the opportunities, several challenges exist to facilitating useful processes. 

\subsection{Inclusion, Diversity, Equity, and Privacy in XR}
\label{lbl_subsec_xr_rw}
Inclusion means providing equal opportunities to benefits and resources regardless of individual differences such as ethnicity, gender, health, or sexual orientation. Facilitating inclusion often promotes diversity and equality. There have been different focus points regarding inclusion and diversity in XR research. For instance, Peck et al.~\cite{8998141} found that recent XR research includes significantly underrepresented women as authors and experiment participants. Later, Peck et al.~\cite{9646525} highlighted differences between underrepresented groups compared to commonly studied populations regarding usability and depicted the lack of generalizability of previous VR research. While engaging the underrepresented populations in XR research and conducting studies with representative samples differ from supporting populations according to individual needs in XR, the latter is needed to personalize the XR experiences and attract underrepresented populations to XR. With this aim, Ajri et al.~\cite{virtual_aivantage_2023} built a VR-based tool that leverages LLMs for interview preparation for underrepresented individuals. With their versatile nature, fine-tuning opportunities, and different prompting techniques, LLMs can provide a good step to achieve this goal. 

To support diverse user populations, it is essential to understand the differences between user groups in XR, especially in how they perceive and behave in XR. To this end, researchers analyzed different user characteristics such as gender~\cite{Gao2023_ijaied, gender_perception_vr_chi17}, health~\cite{sahin2018second, Emmelkamp_2021}, expertise~\cite{hosp_expertise_vr_goalkeeper, gao_ismar_2023}, sexual orientation~\cite{Wenzlaff_2016, Rahman_Koerting_2008}, and race~\cite{own_race_bias_2012, seitz_race_bias_VR_2020}. For instance, Gao et al.~\cite{Gao2023_ijaied} found with explainable machine learning that girls and boys visually behave differently when they attend a lesson in VR. Furthermore, Hosp et al.~\cite{hosp_expertise_vr_goalkeeper} could distinguish different goalkeeper expertise levels in the VR context using eye movements. 

Regarding health-related scenarios, the usability and acceptability of smartglasses have been assessed with children with autism spectrum disorder (ASD) and their caregivers~\cite{sahin2018second}. While the authors found that their audience found the smartglasses acceptable and usable, the reason the study was conducted in the first place is that people with ASD often have issues with social communication and interaction, leading to different ways of learning, moving, or paying attention~\cite{ASD_symptoms}. Similar trends were found in race and sex research as well, with humans remembering faces of their race better than the other faces~\cite{own_race_bias_2012}, and visual behaviors being representative of sexual preferences~\cite{Wenzlaff_2016}. While sensor data revealing different user characteristics are not only limited to XR~\cite{own_race_bias_2012, Wenzlaff_2016} and aforementioned user attributes~\cite{soren_preibusch_2014, Kroeger2020}, previous research indicates that there is no one-size-fits-all approach for personalizing the 3D user interfaces and creating adaptive support for the users considering the distinct behaviors. Customizing LLMs with fine-tuning and prompt engineering can facilitate efficient support for different user populations and profiles. Recent work in the context of AR and LLMs~\cite{hajahmadiarele} also emphasized the potential of personalization with LLMs for language learning. 

Studying different populations and providing personalized experiences might lead to accurately identifying users, especially with machine learning techniques. This identification is convenient for supporting the utility of adaptive interfaces; however, it can also be considered a privacy invasion. To address this, a good chunk of research focused on protecting privacy, especially considering biometric sensor data, such as eye movements~\cite{bozkir2023eyetracked, bozkir_thesis_2022}, with statistical notions such as differential privacy~\cite{diff_privacy_2021_efe} and empirical ways of streaming the sensor data~\cite{9382914}. Other research investigated going incognito in VR by leveraging differential privacy~\cite{nair_going_incognito_2023} and deep motion masking that facilitates real-time anonymization~\cite{nair2023deep} and showed that practical privacy-utility trade-offs are possible. 

Beyond the technical approaches focusing on the conundrum between privacy and utility, it is equally vital to understand the privacy behaviors of users so that human-centered solutions can be designed. Focusing on user privacy concerns about AR glasses, Gallardo et al.~\cite{gallardo_pets2023} found that when different data types and uses are considered, user privacy attitudes and reservations are context-dependent. Denning et al.~\cite{denning_etal_2014} studied user privacy perspectives for AR and indicated participants' requests to be asked before being recorded. Lebeck et al.~\cite{lebeck_etal_2018} showed that bystander privacy is an important concern in a multi-user AR. More recently, O'Hagan et al.~\cite{ohagan_etal_2023} stated the importance of user activity and its relation with bystanders in everyday AR. Considering the research on the technical solutions and usable privacy in XR, it is an open question whether utilizing LLMs with user sensor data in XR will lead to novel privacy leaks and how users' privacy attitudes will be in these situations. The research community has very recently started to emphasize the importance of the need for research on human aspects of privacy issues of LLMs, such as understanding users’ mental models and preferences for privacy controls~\cite{li2024humancentered}. The same need also exists when LLMs are embedded into XR in immersive settings. 

\section{Opportunities, Research Directions, and Challenges}
\begin{figure}[!t]
  \centering
   \includegraphics[width=1\linewidth, keepaspectratio]{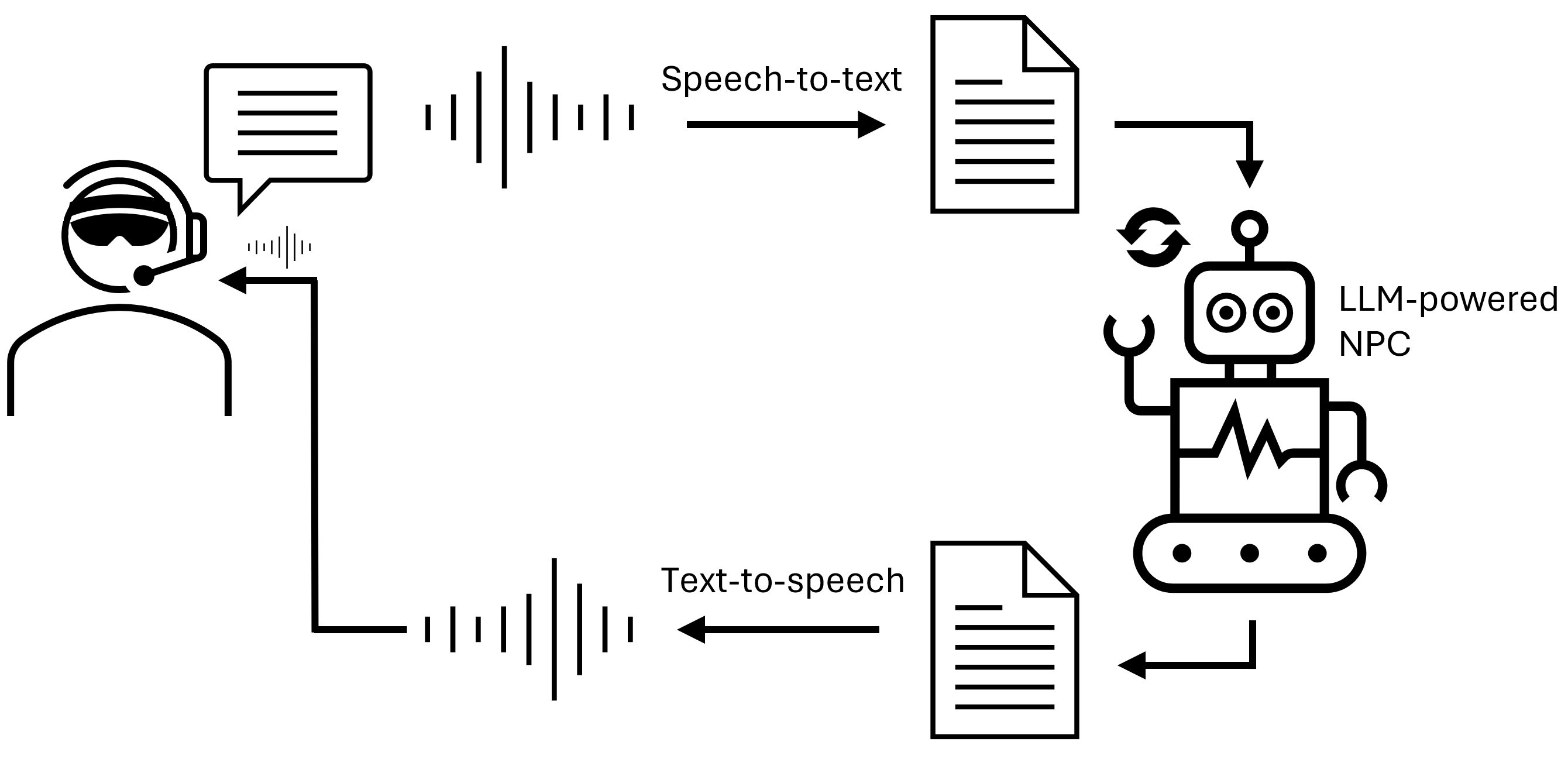}
  \caption{An example of a possible data processing pipeline.}
  \Description{Technical workflow of the proposed processing pipeline. First, audio input from the user in XR is obtained and fed to a speech-to-text model. Then, the output text from this model is sent to an LLM for a response from the LLM-powered virtual avatar in XR. As a last step, the response of the LLM agent is provided to the text-to-speech model to provide the audio response back to the user.}
  \label{fig_arch}%
\end{figure}

Large language models, with their versatile conversation capabilities, fine-tuning possibilities, and prompting techniques, hold an immense upside for XR, particularly if such models are utilized as NPCs or narratives. However, not much research has been conducted to this end. In the very naive scenario, to integrate LLMs in XR, it is essential to utilize speech-to-text and text-to-speech models~\cite{pratap2023scaling} to enable information transition from the user to LLM-powered NPCs and vice versa in an auditory way. Figure~\ref{fig_arch} depicts an example processing pipeline. While there might be some drawbacks to such a pipeline, like latency issues if online services are used for obtaining the responses from LLMs or the requirement of a significant amount of storage in case LLMs are deployed locally within the XR spaces as they can easily occupy several hundreds of gigabytes, these issues can be solved by good engineering practices and deploying customized models on a shared location communicated efficiently via web services. Furthermore, multimodal LLMs~\cite{openai2023gpt4v, Deepmind2023} might utilize such a pipeline end-to-end, mitigating issues like latency. Therefore, considering these issues can be handled commendably, we argue that using LLMs will facilitate inclusion in XR and promote more diverse, equal, and engaging XR spaces. However, we also state that having more inclusive and engaging spaces will lead users to spend more time in XR spaces. As a result, more data and likely more sensitive information will be available during the interaction experience. While more research is needed to validate this, we foresee that such an amount of data combined with the already available sensory data from XR (e.g., eye- and hand-tracking) will lead to novel privacy invasions. Considering all, we provide three sets of opportunities and challenges, including several research opportunities. However, each of those should be evaluated cautiously, and the arguments need to be verified with user studies and empirical data. 

\textbf{Inclusion, diversity, and equity:} We argue that pre-scripted and conventional NPCs require a lot of manual labor to serve different user characteristics in XR. For instance, in a skill training scenario, a novice and an expert will have different needs when communicating with an NPC. Conventionally, two scripts or AI agents must be created to support these two users. Still, such agents might be perceived as fictitious regarding their knowledge or the overall conversation quality due to rule-based conversation generation or agents trained with small datasets for particular uses. In contrast, even the pre-trained LLMs without fine-tuning can provide personalized experiences to users if they are prompted for specific uses~\cite{prompt_engineering_openai}, such as ``When I ask for help, provide a response considering I am an [expert/novice] in XR.'' With fine-tuning, LLMs can even provide more adaptive responses, and we argue that such LLM-powered XR spaces will be more inclusive by motivating and attracting diverse types of users. Additionally, as any user characteristics can be supported in a personalized manner with LLMs, this will facilitate equal opportunities for users. 

Therefore, the major research direction concerns how different user characteristics perceive the LLM-powered XR spaces and whether such spaces can motivate users with different characteristics comparably and facilitate diversity and equality. However, LLMs can also hallucinate~\cite{ye2023cognitive}, and it is a challenge to mitigate the effects of hallucinations. 

\textbf{User engagement:} We think that LLM-powered XR spaces through NPCs or narratives will engage users more than conventional conversational agents due to LLMs' abilities to create human-like responses. As a result, the overall immersion and interactivity within the XR spaces will be enhanced. Consequently, these enhancements will likely captivate users' attention and encourage them to spend more time within XR spaces. The rich and dynamic narratives and realistic interactions facilitated by LLMs would elevate the engagement value of XR and open up new ways for creative storytelling and interaction, transforming how users perceive and behave in XR spaces. 

The major research direction concerning user engagement is understanding whether LLM-powered spaces can significantly increase user engagement compared to conventional XR. When this is the case, users likely provide more information about themselves during their conversations with the NPCs in XR. This might include more sensitive information about themselves, leading to ethical challenges regarding data privacy. 

\textbf{Privacy:} We hypothesize that LLM-powered XR will engage users more, and the users will provide more information about themselves, including the sensitive ones, due to the increased engagement and interaction time. As a result, we expect that such sensitive information obtained during the interaction combined with other sensor data will likely end up with novel privacy invasions about users. It is already known that LLMs have several privacy issues~\cite{smith2023identifying, neel2023privacy}, and a more in-depth investigation should be carried out, as in our case, XR will combine data from LLM interaction and multimodal sensor data. Furthermore, we argue that such privacy invasions should be addressed by also asking users about their privacy concerns and preferences as it has been done in other domains~\cite{naeini2017privacy, felt_etal_2012_smartphones, gallardo_pets2023} by observing whether or not there is a gap between privacy expectations and behaviors~\cite{usable_privacy_xr_vision_2022}. Very recent work also emphasized the usable privacy aspects of conversational agents that are powered by LLMs, and the authors stated that human-like interactions encourage more privacy-sensitive disclosures~\cite{Zhang_etal_CHI24}, as we also argue. However, previous research has not considered the XR aspect, and to this end, a greater number of empirical evaluations are needed to validate privacy-related issues. Therefore, such aspects form opportunities, especially for designing privacy-aware and user-centric XR spaces. 

Therefore, the research direction we identified under the privacy and ethics umbrella concerns understanding whether novel privacy invasions and leaks occur in these novel spaces, understanding users' general privacy attitudes, and designing privacy-enhancing methods when the former is the case. However, since the LLM research is moving very fast compared to others, some aspects might need to be evaluated in a longitudinal manner, which is a challenge. 

\section{Conclusion}
This paper argues for embedding LLMs as NPCs or narratives into XR. We state that due to the versatile conversational capabilities of LLMs, as they are trained with a massive amount of text data from the Internet, prompting techniques, and fine-tuning possibilities, LLM-powered NPCs and narratives will provide significantly improved personalized experiences in XR, regardless of different user characteristics. We argue that these will enhance the motivation to use XR more frequently in an engaged way and facilitate inclusion, leading to more diverse populations using XR. Furthermore, as such personalization can support any user type, LLM-powered XR spaces will also promote equity. Lastly, we underline the importance of privacy and claim that possible novel privacy invasions and users' privacy attitudes should be investigated. 


\bibliographystyle{ACM-Reference-Format}
\bibliography{paper}

\end{document}